\documentclass[12pt]{article}
\usepackage{a4wide,graphicx}
\usepackage{cite}
\usepackage{amssymb}
\usepackage{amsmath}

\begin{document}

\title{$\phi$ meson transparency in nuclei from $\phi N$ resonant interactions}

\author{D.~Cabrera$^1$, A.~N.~Hiller Blin$^2$, M.~J.~Vicente Vacas$^2$, P.~Fern\'andez de C\'ordoba$^1$\\
$^1$ Instituto Universitario de Matem\'atica Pura y Aplicada, \\
Universidad Polit\'ecnica de Valencia, 46022 Valencia, Spain\\ 
$^2$ Instituto de F\'{\i}sica Corpuscular, Universidad de Valencia--CSIC,\\
Institutos de Investigaci\'on, Ap. Correos 22085, E-46071 Valencia, Spain.
}

\maketitle
\begin{abstract}
We investigate the $\phi$ meson nuclear transparency using some recent theoretical developments on the $\phi$ in medium selfenergy. The inclusion of direct resonant
$\phi N$-scattering and the kaon decay mechanisms leads to a $\phi$ width much larger than in most previous theoretical approaches. 
 The model has been confronted with photoproduction data from CLAS and LEPS and the recent proton induced $\phi$ production from COSY finding an overall good agreement. The results support the need of a quite large  direct $\phi N$-scattering contribution to the selfenergy.
\end{abstract}



\section{Introduction}

The light vector meson properties in dense/hot nuclear matter have been intensively studied the last decades in the search, among others, of any signal of chiral symmetry restoration. A good review of the related physics can be found in Refs.~\cite{Rapp:1999ej,Hayano:2008vn}. These mesons are particularly appealing because their dileptonic decays can provide a relatively clean information of the nuclear medium interior as opposed to strong decays undergoing a sizable final state  interaction before the detection of the decay products.
 In addition, the $\phi$ meson is very narrow in vacuum and is well separated from the $\rho$ and the $\omega$ mesons what could help in the experimental analysis and allow for the measurement of any modifications of its mass or width. 

Experimentally, $\phi$ production and its decays, both hadronic and electromagnetic, have been investigated in  heavy ion collisions by the STAR and ALICE collaborations~\cite{Wada:2013mua,Abelev:2014uua}. In cold nuclei, $\phi$ production has been studied at Spring8~\cite{Ishikawa:2004id}, KEK~\cite{Muto:2005za}, Jefferson Lab~\cite{Wood:2010ei} and J\"ulich~\cite{Hartmann:2012ia}.
One of the findings is that, whereas the $\phi$ mass in the medium is scarcely modified if at all, the width is much larger than in vacuum~\cite{Ishikawa:2004id,Muto:2005za,Wood:2010ei,Hartmann:2012ia,Muhlich:2005kf}. Actually, the  in-medium $\phi$ width seems to be substantially larger than predicted by most theoretical models. 
 
This width is expected to come mostly from the decay $\phi\rightarrow K\bar{K}$, which is dominant in vacuum. The medium effects modifying it have been much
studied~\cite{Ko:1992tp,Klingl:1997tm,Oset:2000eg,Cabrera:2002hc,Gubler:2016itj,Cobos-Martinez:2017vtr} and involve a quite rich dynamics. In nuclear  matter, the kaons are just mildly repelled and will move out of the nucleus. However,  antikaons are attracted by the nuclear medium and can also be absorbed leading to hyperons and resonances such as $\Lambda(1405)$ and others. These mechanisms are instrumental leading to a large $\phi$ width. For instance, in Ref.~\cite{Cabrera:2002hc}, we obtain $\Gamma_\phi\approx 30$ MeV at normal nuclear density to be compared to 4 MeV in vacuum. Still, that result is not large enough to describe the experimental data. This failure has been the cause for a search for additional mechanisms which could contribute to the meson decay.  

In Ref.~\cite{Cabrera:2016rnc}, we explored the
$\phi$ selfenergy pieces related to some direct $\phi$-nucleon interaction channels not previously considered. There, $\phi$-nucleon elastic scattering proceeds via $K^*$-hyperon loops which give rise to a selfenergy with real and imaginary parts.  Our work was based in some recent studies analyzing the vector meson scattering with baryons in two different schemes.  Both models account for a relatively strong $\phi$ nucleon interaction. As a consequence of these mechanisms the $\phi$ meson gets an additional broadening up to 40-50 MeV  and a mild attraction at normal nuclear density. Our purpose here is to test the results of the model of Ref.~\cite{Cabrera:2016rnc}  comparing with the available data and check whether a satisfactory description of the $\phi$ selfenergy in cold nuclear matter has been reached. We will focus on its controversial imaginary part, or equivalently the $\phi$ width.

 A direct extraction of the in-medium width via the analysis of the invariant mass of the decay products poses some difficulties. For instance, in Ref.~\cite{Muto:2005za} the dilepton channel was measured in carbon and copper nuclei for 12 GeV $p+A$ reactions.
With this kinematics, most of the $\phi$ mesons move very fast in the forward direction and escape from the nucleus before decaying. As a consequence, the observed width is frequently the free one.
Nonetheless, a clear broadening was observed for the heavier nucleus and when only the slower $\phi$  mesons were selected. On the other hand, the dominant decay channel, $\phi\rightarrow K\bar{K}$, presents some additional challenges related to the final state interaction. The strong antikaon absorption restricts the visibility of decays that happen at high densities far from  the surface. Also the real part of the optical potential, including Coulomb, modifies the kaon trajectories and distorts the invariant mass of the system.
 
Another observable, sensitive to the imaginary part of the $\phi$ selfenergy is the nuclear transparency ratio given by the quotient of the cross sections for $\phi$ production on nuclei and on a free nucleon. This quantity depends on the loss of flux in the medium and thus on the width of the $\phi$ meson and its density dependence. The transparency has been measured in photoproduction by the LEPS and CLAS collaborations~\cite{Ishikawa:2004id,Wood:2010ei}. This process had been suggested in Ref.~\cite{Cabrera:2003wb} and was also studied in Ref.~\cite{Muhlich:2005kf}.
Transparency for the case of proton induced $\phi$ production is more complicated due to the initial state interaction of the proton beam that leads to some secondary production mechanisms such as $p N\rightarrow \pi NN$ followed by $\pi N\rightarrow \phi N$. This process had been studied in Refs.~\cite{Magas:2004eb,Paryev:2008ck,Sibirtsev:2008ib} and has been recently measured at J\"ulich~\cite{Polyanskiy:2010tj,Hartmann:2012ia}. 

In this paper, we present a study of the $\phi$ nuclear transparency for both photon and proton induced production on nuclei using the theoretical model from Ref~\cite{Cabrera:2016rnc}. We will start by giving a brief reminder of the $\phi$ selfenergy model and introduce the formalism used in the calculation. Then we proceed to the comparison with the experimental data.

\section{Theoretical model} 
Two sources of $\phi$ selfenergy in nuclear matter are considered here, the mechanisms related to the $K\bar{K}$ decay, that will be denoted as kaon cloud,  and those coming from $\phi N \rightarrow\phi N $ resonant scattering mediated by hyperon + vector meson and other intermediate coupled channels. 

In vacuum, the largest decay channel (83\%) is $\phi\rightarrow K\bar{K}$. At leading order, the  $\phi$ selfenergy is obtained by evaluating $K(\bar{K})$ loop and tadpole diagrams. The nuclear medium effects are incorporated by properly dressing the kaon and antikaon propagators with their selfenergies originating from the $KN(\bar{K}N)$ $s$ and $p$-wave interaction. Details on the calculation of this contribution to the $\phi$ selfenergy can be found in Ref.~\cite{Cabrera:2002hc} and for the kaon/antikaon selfenergy we use the results from Refs.~\cite{Ramos:1999ku,Tolos:2006ny}.

The $K$ selfenergy is relatively simple. The $KN$ amplitude is elastic and given the absence of resonances depends very slowly on the energy. To a good approximation the selfenergy can be cast in the $T\rho$ form.
The  $\bar{K}$ case is more involved. The $p$-wave part of the selfenergy includes the coupling to several particle-hole excitations such as $\Lambda(1115)N^{-1}$, $\Sigma(1195)N^{-1}$ and $\Sigma^*(1385)N^{-1}$. The $s$-wave part of the selfenergy is calculated in a unitarized chiral model and is dominated by the excitation of the $\Lambda(1405)$ resonance. A specially careful and self-consistent treatment of the many-body corrections is required in this case because of the vicinity to the $\bar{K}N$ threshold. As a result, a quite large width is obtained for the antikaons. Furthermore,  the real part of the optical potential shows an attraction of $-60$ MeV at normal nuclear matter density for antikaons at rest in contrast to the mild repulsion in the kaon case. 

 The novelty of Ref.~\cite{Cabrera:2016rnc} was the calculation of the contribution to the $\phi$ selfenergy in the medium related to the $\phi N$ elastic scattering amplitude. We relied upon the results of two different schemes recently developed to describe the vector meson--baryon scattering. The first one~\cite{Oset:2009vf,Oset:2012ap,Ramos:2013wua}  obtains the low-energy vector meson--baryon amplitude within the hidden local symmetry (HLS) approach. The second one~\cite{Gamermann:2011mq} uses an SU(6) spin-flavor symmetry extension of the SU(3) chiral perturbation theory Lagrangian. This leads to the generalization of the Weinberg--Tomozawa interaction between pseudoscalar and vector mesons, and baryons from the light octet and decuplet. In both schemes the scattering amplitude is calculated in a coupled channels unitarized approach. These models have been successful in reproducing  masses and decay widths of some negative parity resonances and the HLS one has also been tested and constrained in the analysis of the $\gamma p\rightarrow K\Sigma$ reaction~\cite{Ramos:2013wua}. At the lowest order, in these models, there is no direct
$\phi N \rightarrow \phi N$ interaction but that   process happens via loops such as 
$\phi N \rightarrow K^* \Lambda \rightarrow\phi N $. These loops, on the other hand, produce an imaginary part for the scattering amplitude through the opening of some decay channels.

The contribution to the selfenergy is then obtained by summing the scattering amplitude over the initial nucleon Fermi distribution. Also  Pauli blocking is taken into account by replacing the vacuum nucleon propagators that appear in the calculation by single-particle propagators in the Fermi gas approximation. The new mechanisms produce a moderate momentum dependence of the $\phi$ selfenergy reflecting the presence of some resonances on the $\phi N$ amplitude.  Furthermore, the predictions of the two theoretical models differ significantly at low momenta for both real and imaginary parts of the optical potentials. Close to threshold the attraction  ranges from 5 to 40 MeV, what could strongly affect the existence and spectrum of possible $\phi$ meson nuclear bound states~\cite{Cobos-Martinez:2017woo}. The imaginary part is stronger for the SU(6) model, though both models provide a larger contribution than the mechanisms related to the $K\bar{K}$ decay.

\subsection{Nuclear transparency: photoproduction}

We start discussing the case of $\phi$ nuclear photoproduction reactions. In this case shadowing is negligible. Thus, the reaction takes place in the whole nucleus and the cross section can be approximated by
\begin{equation}
\label{eq:1}
\frac{d\sigma_A}{d\Omega}=\int d^3r\,\rho(r) \frac{d\sigma_N}{d\Omega} F_{ABS},
\end{equation}
where $\frac{d\sigma_N}{d\Omega}$ and $\frac{d\sigma_A}{d\Omega}$ are the elementary-nucleon and nuclear differential cross section, respectively.  $F_{ABS}$ is an absorption factor accounting for the $\phi$ meson lost flux on its way out of the nucleus. Here, in the production itself, Fermi motion and Pauli blocking have not been considered. If we also set $ F_{ABS}=1$, omitting  $\phi$ absorption in the nucleus, we would get the trivial result  $\frac{d\sigma_A}{d\Omega}=A\frac{d\sigma_N}{d\Omega}$ where $A$ is the number of nucleons\footnote{ 
Notice the implicit assumption, supported by the experiment~\cite{Chang:2009yq},  that the $\phi$ production cross section from protons and neutrons is very similar.}.

 On the other hand, for energies close to threshold, just for kinematic reasons, the $\phi$ meson goes forwards and is quite fast. The high momentum means that changes of trajectory because of the small real part of the optical potential can be neglected. Also, the quasielastic collisions are very improbable, as the imaginary part of the selfenergy is fully dominated by inelastic channels according to our theoretical models. Therefore, to a good approximation, the $\phi$ meson will move forward until it gets out of the nucleus or it is absorbed. Thus, we can model the absorption factor in an eikonal form as~\cite{Cabrera:2003wb}
\begin{equation}\label{eq:2}
F_{ABS}=\exp\left(-\int_{0}^{\infty}dl\frac{1}{p}\, \text{Im}\,\Pi(p,\rho(r'))\right),
\end{equation}
where $\Pi(p,\rho(r))$ is the $\phi$ selfenergy as a function of its momentum $p$ and at the nuclear  density $\rho$, $\vec{r}$ is the $\phi$ production point. Finally,   $\vec{r'}=\vec{r}+l \,\vec{p}/|\vec{p}|$.   
  As long as the integrand of Eq.~\ref{eq:1} does not depend on the direction of the $\phi$ momentum, other than via $\frac{d\sigma_N}{d\Omega}$, we can write the following ratio between the nuclear and the nucleon cross section
\begin{equation}
\label{eq:3}
P_{out}\equiv\frac{\sigma_A}{A\,\sigma_N}=\frac{1}{A}\int d^3r\,\rho(r)\,\exp\left(-\int_{0}^{\infty}dl\frac{1}{p}\, \text{Im}\,\Pi(p,\rho(r'))\right),
\end{equation}
 which measures the transparency of the nucleus to the $\phi$ meson.

\begin{figure}[ht]
\centering
\includegraphics[width=0.75\textwidth]{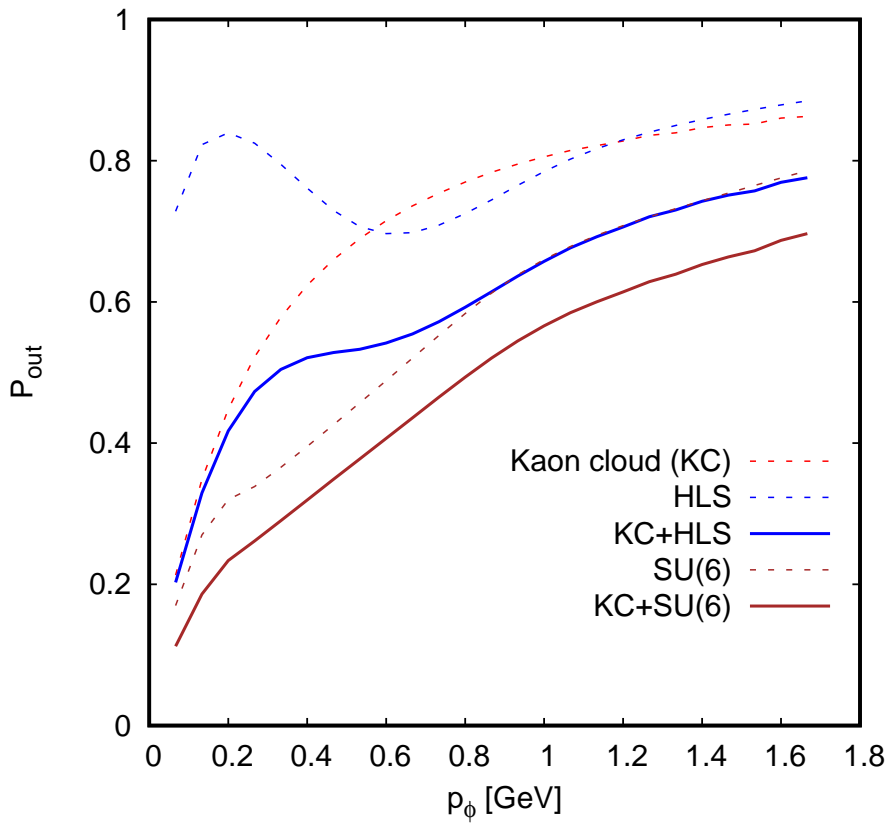} 
\caption{Transparencies for the nucleus $^{20}$Ne as a function of the $\phi$ momentum with kaon cloud selfenergy only or adding the $\phi$ selfenergies from the HLS~\cite{Oset:2009vf,Oset:2012ap} or the SU(6) model~\cite{Gamermann:2011mq}.}
\label{fig:1}
\end{figure}

The effect on the transparency observable for the $\phi N$ resonant scattering is substantial, as expected from its  large contribution to the $\phi$ selfenergy~\cite{Cabrera:2016rnc}.
In Fig. \ref{fig:1}, we show this ratio between  cross sections for $^{20}$Ne as a function of the $\phi$ momentum for the theoretical models considered in this paper.
The nuclear density profiles for all cases  have been taken from~\cite{DeJager:1987qc,DeJager:1974liz}. The inclusion of the new $\phi N$ scattering mechanisms leads to a much stronger absorption for the whole momentum range explored than the kaon cloud alone. Additionally, the HLS model shows a strong energy dependence at relatively low ($<600$ MeV) momenta. At higher momenta the nuclear transparency increases for all cases.

  The only nuclear effects considered in this result and in Eq.~\ref{eq:3} are those related to $\phi$ absorption,  incorporated into the calculation of $\Pi$, the $\phi$ selfenergy. Other nuclear effects affecting the production mechanism, rather than the $\phi$ propagation, are the
Fermi motion of the initial nucleon and the Pauli blocking of the final one on the $\gamma N\rightarrow \phi N$ process. Pauli blocking will imply a reduction of the $\phi$ production cross section.
The Fermi motion will distort the distribution of the final meson and nucleon and affect the Pauli blocking itself. The flux reduction due to these sources can be estimated for photon induced reactions by including in the integrand of Eq.~\ref{eq:3} a factor considering a Fermi average of these effects~\cite{Cabrera:2003wb,GarciaRecio:1987ik}:
\begin{equation}
G(Q,\rho)=1-\Theta(2-\tilde{Q})\left( 1-\frac{3}{4} \tilde{Q}+
\frac{1}{16} \tilde{Q}^3
  \right),
\end{equation}
where $\tilde{Q} = |\vec{Q}|/k_F$, $\vec{Q}$ is the momentum transfer and $k_F$
is the Fermi momentum of the nucleons.
\begin{figure}[ht]
\centering
\includegraphics[width=0.75\textwidth]{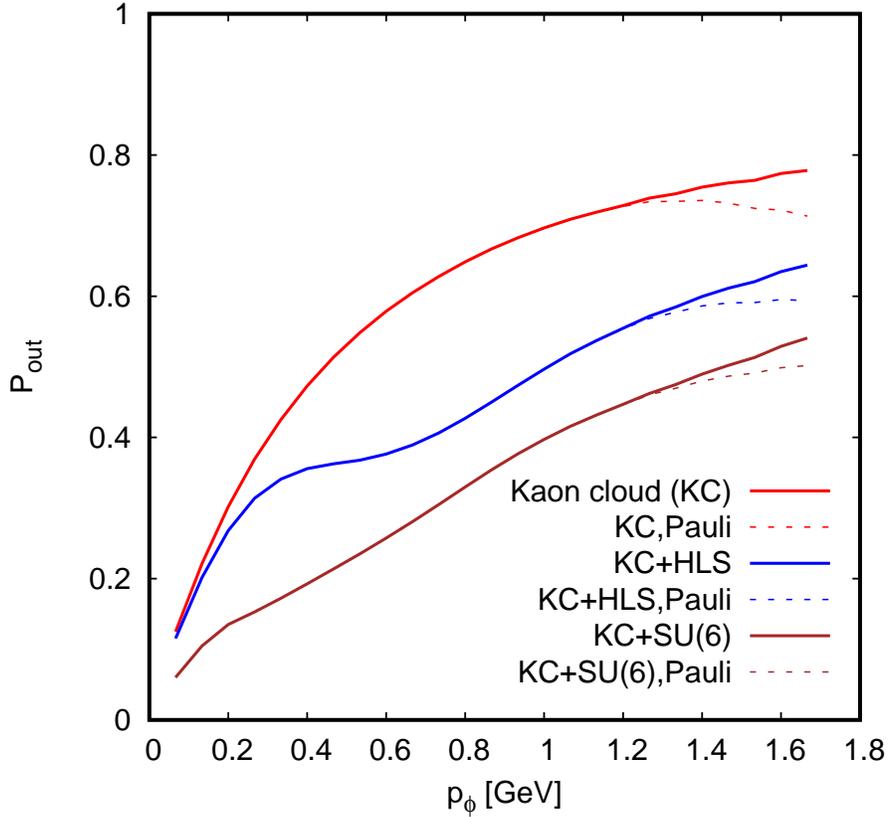} 
\caption{Transparencies as a function of  $\phi$ momentum for photoproduction processes, with ($\theta_\phi=0$ degrees) and without Pauli blocking for $^{64}$Cu.}
\label{fig:2}
\end{figure}
In Fig.~\ref{fig:2}, we show how the transparency is modified by the Pauli blocking of the final nucleon. The result depends on the scattering angle. For the figure we have selected forward $\phi$ scattering that maximizes the change. Opening the angle increases the transfer momentum and as soon as it is above $2k_F$, Pauli blocking becomes ineffective.  There is a small reduction at high $\phi$ momentum and practically no change below 1.2 GeV. This reduction will also affect transparency ratios comparing different nuclei because of the variation of the average density, and thus of the Fermi momentum. However, the dependence of the Pauli blocking correction on the nuclear size, beyond $A\approx 10$, is minimal as shown in Ref.~\cite{Cabrera:2003wb}. 

\begin{figure}[ht]
\centering
\includegraphics[width=0.75\textwidth]{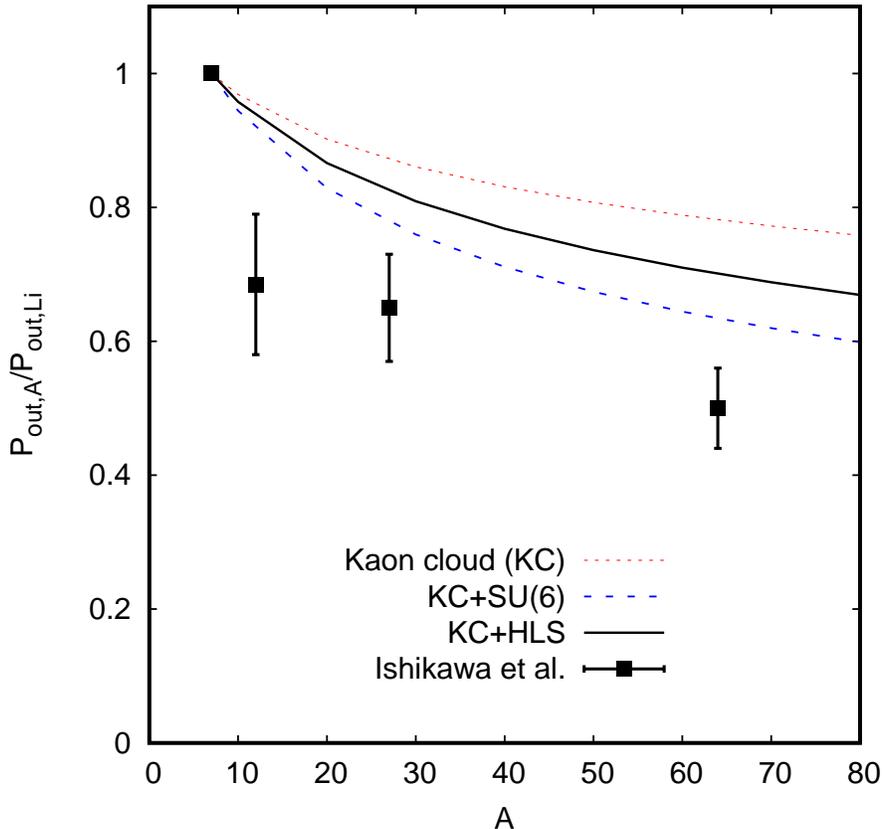} 
\caption{Ratio of $\phi$ photoproduction transparencies as a function of the atomic number compared with data from LEPS~\cite{Ishikawa:2004id}.}
\label{fig:3}
\end{figure}

In Fig.~\ref{fig:3}, we compare our model with data from LEPS~\cite{Ishikawa:2004id} which measured the transparency detecting the $\phi$ mesons through  their $K\bar{K}$ decay.
The transparencies are normalized to that of lithium, the lightest nucleus measured in the experiment. In this way, some systematic errors could be reduced. Our presented results are obtained assuming forward scattering, thus maximizing the Pauli blocking effects. Removing Pauli blocking would push up by less than a 5\%~\cite{Ishikawa:2004id} the three curves.
The photon spectrum had  energies ranging  from 1.5 to 2.4 GeV. We take an average momentum, $p_\phi=1.8$ GeV as suggested in ~\cite{Ishikawa:2004id}. We find that the inclusion of the $\phi N$ scattering mechanisms improves the agreement for both models. In principle, the largest absorption corresponding to the SU(6) model is favored. However, we find that it is very hard to reproduce the steep change in data from lithium to carbon, even when artificially increasing the absorption by a large factor.

\begin{figure}[ht]
\centering
\includegraphics[width=0.75\textwidth]{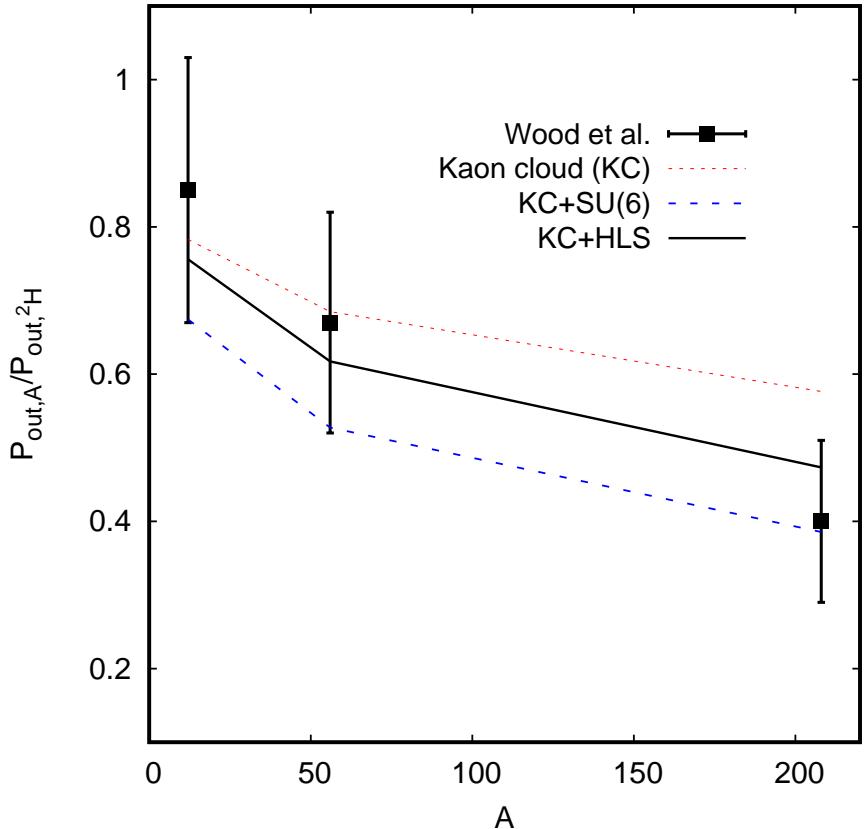} 
\caption{Transparency ratios for $\phi$ photoproduction as a function of the atomic number compared with data from CLAS~\cite{Wood:2010ei}. Curves to guide the eye.}
\label{fig:4}
\end{figure}

In fig.~\ref{fig:4}, we compare our model with data from CLAS~\cite{Wood:2010ei}. In this case, the $\phi$ meson was detected via the $e^+e^-$ decay avoiding the complication of the final state kaon interactions and some other difficulties in the background subtraction and the experimental analysis. The average $\phi$ momentum is 2 GeV, only slightly larger than for LEPS. According to our model the transparencies at such close energies should be similar. The results here are consistent with data. Lead transparency, the one with the strongest nuclear effects, favors the inclusion of the new mechanisms and overall the best fit corresponds to the HLS model. Nonetheless, the large uncertainties prevent us from reaching strong conclusions.  The much larger $\sigma_{\phi N}$ cross section that would be required to accommodate LEPS data would spoil the agreement with CLAS. Thus, the two available photoproduction experimental results seem hardly compatible.

\subsection{Nuclear transparency: proton induced production}

The theoretical description of proton induced $\phi$ production is more complicated~\cite{Magas:2004eb,Paryev:2008ck,Sibirtsev:2008ib} even when assuming that the quasifree mechanism $pN\rightarrow pN\phi$ is dominant. First, we must consider the initial state interaction of the proton. A simple approximation is to include an additional factor to account for the proton flux reduction,
\begin{equation}\label{eq:5}
F_{IN}=\exp\left(-\int^{\vec{r}}_{\infty}\sigma_{NN} \;\rho(r') \;dl \right),
\end{equation}
where $\sigma_{NN}$ is the full nucleon-nucleon cross section. 
From here on, we adapt to the COSY/ANKE setup of Ref.~\cite{Hartmann:2012ia}. There, the protons have a kinetic energy of 2.83 GeV. It is  close to the reaction threshold and thus Pauli blocking is irrelevant for the primary reaction $NN\rightarrow NN\phi$ because the final nucleons have a too large momentum. On the other hand, for the initial distortion both $\sigma_{pn}$ and $\sigma_{pp}$ are around 42~mb~\cite{Olive:2016xmw}.
A second change with respect to the photoproduction process  is the sizable isospin asymmetry in the production cross section. 
According to both experimental data and theoretical models~\cite{Maeda:2006wv,Titov:1997kt,Kaptari:2004sd} the cross section for $pn\rightarrow pn\phi$ is substantially larger than for $pp\rightarrow pp\phi$.
Also, the  $pn\rightarrow d\phi$ process, which further enhances the relevance of neutrons, is of comparable size~\cite{Maeda:2006wv}.
This isospin asymmetry is taken into account substituting $\sigma_N$ in Eq.~\ref{eq:1} by 
\begin{equation}\label{eq:6}
\left\{ N (\sigma_{pn\rightarrow pn\phi}+\sigma_{pn\rightarrow d\phi})+Z\sigma_{pp\rightarrow pp\phi}\right\} /A,
\end{equation}
with $Z$ and $N$ the number of protons and neutrons and $A=N+Z$.
We use for these cross sections the parametrizations from Ref.~\cite{Paryev:2008ck}. Obviously, this isospin asymmetry leads to a relatively larger $\phi$ production for heavier nuclei which have more neutrons than protons. The effect is of the order of 10\% for lead at the energy of Ref.~\cite{Hartmann:2012ia}. 

\begin{figure}[ht]
\centering
\includegraphics[width=0.75\textwidth]{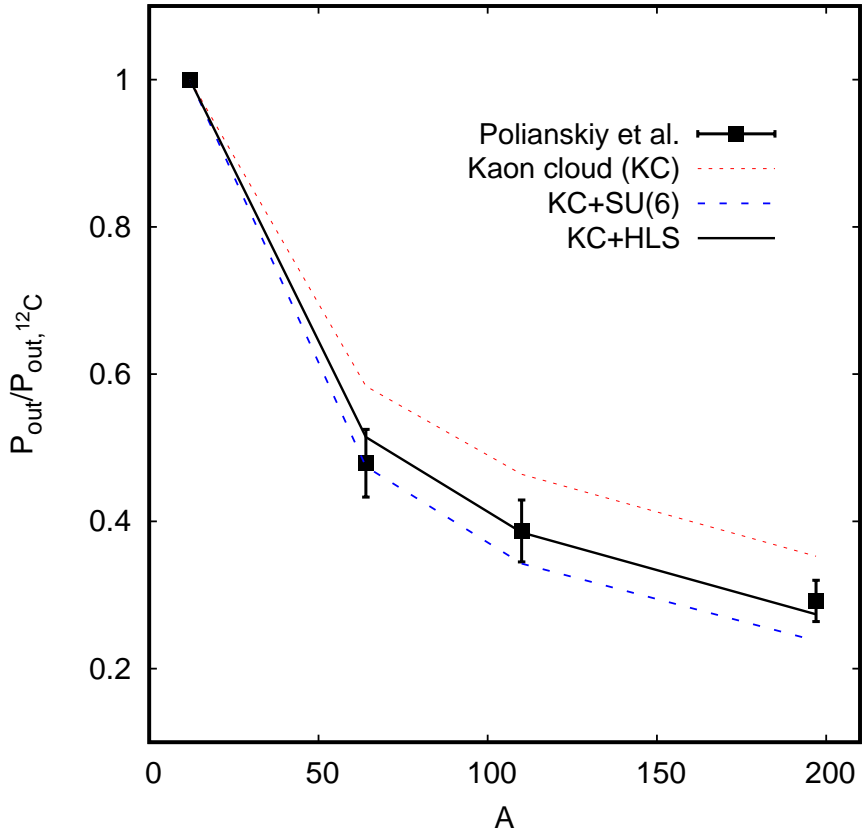}\caption{
Nuclear transparency ratios for the $pA\rightarrow \phi X$ reaction 
compared to data from Ref.~\cite{Polyanskiy:2010tj}.}
\label{fig:5}
\end{figure}
Including the shadowing factor of Eq.~\ref{eq:5} and the isospin correction from Eq.~\ref{eq:6} we compare our results with data from Ref.~\cite{Polyanskiy:2010tj} in Fig.~\ref{fig:5}. In the calculation we have taken an average $\phi$ momentum of 1.3 GeV, which approximately corresponds to the experimental peak of the $\phi$ production differential cross section for all nuclei~\cite{Hartmann:2012ia} and also of the phase space distribution of the elementary $NN\rightarrow NN\phi$ process at the studied energy.

The agreement is fair for all models and a simple $\chi$-squared analysis favors the HLS one. We should mention that in  the proton induced process, a good part of the cross section reduction in nuclei comes from the initial state interaction of the proton. Thus, the process is more peripheral and there is less sensitivity to the $\phi$ meson absorption than in photoproduction~\cite{Magas:2004eb}. 

Additionally, there are some caveats to be considered before giving too much weight to these results. The contribution of multistep processes to
the $\phi$ production mechanism could also be important. For instance, the initial nucleon could undergo a quasielastic scattering loosing some energy, followed by $\phi$ production in a second step. Another possibility is the excitation of a $\Delta$ resonance followed by a $\Delta N\rightarrow NN\phi$ process. These two mechanisms were investigated in Ref.~\cite{Magas:2004eb} finding that they were relevant modifying the nuclear cross sections, but hardly affected ratios such as that of Fig.~\ref{fig:5}.  A third mechanism, $\pi$ production followed by $\pi N\rightarrow N\phi$, has been studied in Ref.~\cite{Paryev:2008ck} leading to some enhancement of the nuclear transparency ratios. Given the influence of all these mechanisms, with their large uncertainties, and the smaller sensitivity to the $\phi$ meson absorption, we find that proton induced production is less adequate than $\phi$ photoproduction to obtain information on the $\phi$ selfenergy in nuclear matter.

\section{Conclusions}

We have investigated the $\phi$ meson nuclear transparency using the $\phi$ selfenergy model developed in Ref.~\cite{Cabrera:2016rnc}.  This selfenergy includes direct $\phi N$-scattering mechanisms, evaluated in two different theoretical approaches, in addition to the terms due to the supposedly dominant kaon-cloud interactions. We find that the contribution associated to  $\phi N$-scattering is stronger than assumed in many previous theoretical calculations. 
With this selfenergy, we reproduce well the nuclear transparency data obtained from $\phi$ photoproduction reactions at CLAS. Furthermore, the agreement with the LEPS photoproduction 
data is clearly improved when the $\phi N$-scattering effects are considered. However, an even stronger $\phi$ absorption would be required in this case.
 We find that  CLAS and LEPS data are hardly reconcilable, since they seem to point to different in-medium $\phi$ absorption magnitudes. 

The results also show a good reproduction of the proton induced transparency data. However, this case is less sensitive to the $\phi$ meson properties in the nuclear medium. Namely, large changes of the selfenergy lead to small changes of the transparency which is dominated by shadowing effects. Furthermore, the theoretical modelling is necessarily more involved because of the importance of multistep production mechanisms. 

This work supports the relevance of the direct  $\phi N$-scattering mechanisms on the description of the $\phi$ meson width in the nuclear medium. However, there are still substantial uncertainties in the available theoretical models describing $\phi N$ scattering. This calls for new, more precise experiments, which could help discriminating and constraining  those theoretical models. In particular, the measurement of other observables, such as the spectrum of $\phi$ nuclear bound states, if they exist,  would be instrumental to determine both the real and the imaginary part of the $\phi$ selfenergy in nuclear matter.

\section*{Acknowledgements}
We acknowledge 
A. Ramos, E. Oset, L. Tol\'os and J. Nieves for fruitful discussions and for providing us with numerical  codes implementing their models.
This research has been partially supported by the Spanish Ministerio de Econom\'{\i}a y Competitividad (MINECO) and the European fund for regional development (FEDER) under Contracts FIS2014-51948-C2-2-P and SEV-2014-0398 and by Generalitat Valenciana under Contract PROMETEOII/2014/0068.


\begin{thebibliography}{999}

\bibitem{Rapp:1999ej} 
  R.~Rapp and J.~Wambach,
  Adv.\ Nucl.\ Phys.\  {\bf 25}, 1 (2000)
  [hep-ph/9909229].

\bibitem{Hayano:2008vn} 
  R.~S.~Hayano and T.~Hatsuda,
  Rev.\ Mod.\ Phys.\  {\bf 82}, 2949 (2010).

\bibitem{Wada:2013mua}
  M.~Wada [STAR Collaboration],
  Nucl.\ Phys.\ A {\bf 904-905} (2013) 1019c.

\bibitem{Abelev:2014uua}
  B.~B.~Abelev {\it et al.} [ALICE Collaboration],
  Phys.\ Rev.\ C {\bf 91} (2015) 024609.

\bibitem{Ishikawa:2004id}
  T.~Ishikawa, D.~S.~Ahn, J.~K.~Ahn, H.~Akimune, W.~C.~Chang, S.~Date, H.~Fujimura and M.~Fujiwara {\it et al.},
  Phys.\ Lett.\ B {\bf 608} (2005) 215.

\bibitem{Muto:2005za}
  R.~Muto {\it et al.}  [KEK-PS-E325 Collaboration],
  Phys.\ Rev.\ Lett.\  {\bf 98} (2007) 042501.

\bibitem{Wood:2010ei}
  M.~H.~Wood {\it et al.}  [CLAS Collaboration],
  Phys.\ Rev.\ Lett.\  {\bf 105} (2010) 112301.

\bibitem{Hartmann:2012ia}
  M.~Hartmann, Y.~T.~Kiselev, A.~Polyanskiy, E.~Y.~Paryev, M.~Buscher, D.~Chiladze, S.~Dymov and A.~Dzyuba {\it et al.},
  Phys.\ Rev.\ C {\bf 85} (2012) 035206.
  
\bibitem{Muhlich:2005kf}
   P.~Muhlich and U.~Mosel,
   Nucl.\ Phys.\ A {\bf 765}, 188 (2006).

\bibitem{Ko:1992tp}
  C.~M.~Ko, P.~Levai, X.~J.~Qiu and C.~T.~Li,
  Phys.\ Rev.\ C {\bf 45} (1992) 1400.
  
\bibitem{Klingl:1997tm}
  F.~Klingl, T.~Waas and W.~Weise,
  Phys.\ Lett.\ B {\bf 431} (1998) 254.
  
\bibitem{Oset:2000eg}
  E.~Oset and A.~Ramos,
  Nucl.\ Phys.\ A {\bf 679} (2001) 616.
  
\bibitem{Cabrera:2002hc}
  D.~Cabrera and M.~J.~Vicente Vacas,
  Phys.\ Rev.\ C {\bf 67} (2003) 045203.

\bibitem{Gubler:2016itj}
   P.~Gubler and W.~Weise,
   Nucl.\ Phys.\ A {\bf 954}, 125 (2016).


\bibitem{Cobos-Martinez:2017vtr} 
  J.~J.~Cobos-Mart\'inez, K.~Tsushima, G.~Krein and A.~W.~Thomas,
  arXiv:1703.05367 [nucl-th].

\bibitem{Cabrera:2016rnc} 
  D.~Cabrera, A.~N.~Hiller Blin and M.~J.~Vicente Vacas,
  Phys.\ Rev.\ C {\bf 95}, no. 1, 015201 (2017).

\bibitem{Cabrera:2003wb}
  D.~Cabrera, L.~Roca, E.~Oset, H.~Toki and M.~J.~Vicente Vacas,
  Nucl.\ Phys.\ A {\bf 733} (2004) 130.
  
\bibitem{Magas:2004eb}
  V.~K.~Magas, L.~Roca and E.~Oset,
  Phys.\ Rev.\ C {\bf 71} (2005) 065202.

\bibitem{Paryev:2008ck}
  E.~Y.~Paryev,
  J.\ Phys.\ G {\bf 36} (2009) 015103.

\bibitem{Sibirtsev:2008ib}
  A.~Sibirtsev, H.-W.~Hammer and U.-G.~Meissner,
  Eur.\ Phys.\ J.\ A {\bf 37} (2008) 287.

\bibitem{Polyanskiy:2010tj}
  A.~Polyanskiy {\it et al.},
  Phys.\ Lett.\ B {\bf 695} (2011) 74.

\bibitem{Ramos:1999ku} 
  A.~Ramos and E.~Oset,
  Nucl.\ Phys.\ A {\bf 671}, 481 (2000).

\bibitem{Tolos:2006ny} 
  L.~Tolos, A.~Ramos and E.~Oset,
  Phys.\ Rev.\ C {\bf 74}, 015203 (2006).

\bibitem{Oset:2009vf}
  E.~Oset and A.~Ramos,
  Eur.\ Phys.\ J.\ A {\bf 44} (2010) 445.

\bibitem{Oset:2012ap}
  E.~Oset, A.~Ramos, E.~J.~Garzon, R.~Molina, L.~Tolos, C.~W.~Xiao, J.~J.~Wu and B.~S.~Zou,
  Int.\ J.\ Mod.\ Phys.\ E {\bf 21} (2012) 1230011.

\bibitem{Ramos:2013wua}
  A.~Ramos and E.~Oset,
  Phys.\ Lett.\ B {\bf 727} (2013) 287.

\bibitem{Gamermann:2011mq}
  D.~Gamermann, C.~Garcia-Recio, J.~Nieves and L.~L.~Salcedo,
  Phys.\ Rev.\ D {\bf 84} (2011) 056017.

\bibitem{Cobos-Martinez:2017woo}
  J.~J.~Cobos-Mart\'inez, K.~Tsushima, G.~Krein and A.~W.~Thomas,
  arXiv:1705.06653 [nucl-th].

\bibitem{Chang:2009yq} 
  W.~C.~Chang {\it et al.} [LEPS Collaboration],
  Phys.\ Lett.\ B {\bf 684}, 6 (2010)

\bibitem{DeJager:1987qc}
  H.~De Vries, C.~W.~De Jager and C.~De Vries,
  Atom.\ Data Nucl.\ Data Tabl.\  {\bf 36} (1987) 495.

\bibitem{DeJager:1974liz}
  C.~W.~De Jager, H.~De Vries and C.~De Vries,
  Atom.\ Data Nucl.\ Data Tabl.\  {\bf 14} (1974) 479.

\bibitem{GarciaRecio:1987ik} 
  C.~Garcia-Recio, E.~Oset and L.~L.~Salcedo,
  Phys.\ Rev.\ C {\bf 37}, 194 (1988).




\bibitem{Olive:2016xmw}
  C.~Patrignani {\it et al.} [Particle Data Group],
  Chin.\ Phys.\ C {\bf 40} (2016) no.10,  100001.

\bibitem{Maeda:2006wv}
  Y.~Maeda {\it et al.},
  Phys.\ Rev.\ Lett.\  {\bf 97} (2006) 142301.

\bibitem{Titov:1997kt}
  A.~I.~Titov, B.~Kampfer and V.~V.~Shklyar,
  Phys.\ Rev.\ C {\bf 59} (1999) 999.

\bibitem{Kaptari:2004sd}
  L.~P.~Kaptari and B.~Kampfer,
  Eur.\ Phys.\ J.\ A {\bf 23} (2005) 291.




\end{thebibliography}
\end{document}